# DISCOVERY OF CIV EMISSION FILAMENTS IN M87[1]

Sparks, W.B.[2,3], Pringle, J.E.[3,4], Donahue, M.[5], Carswell, R.[4], Voit, M.[5], Cracraft, M.[3], Martin, R.G.[4]

[3] Space Telescope Science Institute, 3700 San Martin Drive, Baltimore, MD 21218, USA.
[4] Institute of Astronomy, University of Cambridge, Madingley Road, Cambridge, UK.
[5] Physics and Astronomy Department, Michigan State University, East Lansing, MI 48824, USA.

ABSTRACT

Gas at intermediate temperature between the hot X-ray emitting coronal gas in galaxies at the centers of galaxy clusters, and the much cooler optical line emitting filaments, yields information on transport processes and plausible scenarios for the relationship between X-ray cool cores and other galactic phenomena such as mergers or the onset of an active galactic nucleus. Hitherto, detection of intermediate temperature gas has proven elusive. Here, we present FUV imaging of the "low excitation" emission filaments of M87 and show strong evidence for the presence of CIV 1549 Å emission which arises in gas at temperature $\sim 10^5$K co-located with H$\alpha$+[NII] emission from cooler $\sim 10^4$K gas. We infer that the hot and cool phases are in thermal communication, and show that quantitatively the emission strength is consistent with thermal conduction, which in turn may account for many of the observed characteristics of cool core galaxy clusters.

*KEY WORDS*: conduction – galaxies: individual (M87) – galaxies: ISM

1. INTRODUCTION
The close correlation between 'cool core clusters' viewed in X-rays, and optical emission-line nebulae in clusters of galaxies has been recognized for many years, but the physical reason for this connection remains unclear. Optical emission line filamentary structures have been seen and analyzed in many cool core clusters (e.g. Sparks et al., 1989; Voit & Donahue, 1997; Crawford et al., 2005; Sparks et al., 2004; Conselice, Gallagher & Wyse, 2001; Hatch et al., 2006 and references therein). Suggested formation scenarios for the "optical" filaments have included (i) condensations in cooling intracluster medium (e.g. Fabian & Nulsen, 1977; Cowie, Fabian & Nulsen, 1980; Fabian, Nulsen & Canizares, 1984; Heckman et al., 1989; Donahue & Voit, 1991; Revaz, Combes & Salomé, 2007), (ii) already cold interstellar material originating in galaxies falling into the cluster core (e.g. Rubin et al., 1977; Sparks et al., 1989; Sparks, 1992; Braine et al., 1995; Yagi et al., 2007) and (iii) already cold material originating in the central cluster galaxy (e.g. Fabian et al., 2003; Crawford et al., 2005).

Many different excitation mechanisms for the filaments have been considered. For NGC 1275 (Sabra et al. 2000; Conselice, Gallagher & Wyse 2001) and M87 (Sabra et al. 2003)

---

photoionization by the central AGN and by the intracluster medium, by shocks, and by hot, young stars were all considered. These authors concluded that "neither shocks nor photoionization alone can reproduce the emission line intensity ratios" and that some additional source of heating must be present. A study of the optical line ratios in Abell 2597 led Voit & Donahue (1997) to rule out shocks as an excitation mechanism, and to conclude that although hot stars might be the best candidate for producing the ionization, even the hottest stars could not power a nebula as hot as the one observed and that some other non-ionizing source of heat must contribute a comparable amount of power. Similar conclusions were reached by Hatch et al. (2007) for a number of cool-core clusters.

One obvious source of such extra heating comes from the fact that the relatively cool H$\alpha$+[NII] emitting gas is immersed a much hotter surrounding X-ray emitting medium. Thus (Sparks et al. 1989) thermal conductivity is an obvious candidate. Since these two components of the interstellar medium are in thermal contact, there must be a range of gas temperatures in an interface region between the hot coronal ~$10^7$K gas and cooler ~$10^4$K gas. At intermediate temperatures, a major coolant is the strong CIV 1549 Å resonance line. This emission line arises from gas at 50,000K < T < 180,000K. If thermal conduction is effective, there has to be heat flow and a distribution of temperature in the vicinity of the filaments (Sparks 1992, Nipoti & Binney 2004).

M87 (NGC4486) at the center of the Virgo cluster, shows a well-known optical filament system wrapped around the inner radio lobes, in low velocity outflow from the nucleus close to the center, and in modest infall at larger distances. The filaments, like others of this class, are dusty with the dust showing normal extinction characteristics (Sparks, Ford & Kinney 1993; Sabra et al. 2003). The optical filaments are closely related to X-ray filaments, indicative of a physical connection between the low and high temperature gas (Sparks et al. 2004), hence the spatial distribution and strength of intermediate temperature material will constrain plausible mechanisms and transport processes operating within this environment. To test these ideas and a particular thermal conduction model of the filament and excitation, we designed an experiment to seek CIV FUV emission associated with the filaments. A distance to M87 of 16.7 Mpc is assumed (Blakeslee et al. 2009).

2. OBSERVATIONS AND RESULTS

The Hubble Space Telescope (HST) Cycle 17 General Observer proposal GO-11861 used the Advanced Camera for Surveys Solar Blind Channel (ACS/SBC) to obtain FUV images of four cool-core galaxy clusters: two locations within M87 and one location for each of three other clusters known to contain optical line emission. Our strategy is to use two SBC long-pass filters in tandem, which allows us to separate line and continuum emission. The difference of F150LP and F165LP provides a clean "on minus off" CIV image, see Figure 1. We crafted the observations to provide detection limits for CIV that would be 1/5 the strength of H$\alpha$+[NII] in the filaments of M87. Each pointing uses only a single orbit; here we describe the observations of the "outer field" of M87 which were obtained 2009 February 27 in the form of an exposure of 1630 sec duration (jb4g02010) using the F150LP filter, and 870 sec (jb4g02020) with the F165LP filter. The choice of F165LP to complement F150LP provides a stringent guard against potential redleaks that have been found to be present in the ACS/SBC.

Figure 2 shows the F150LP and F165LP count rate images displayed in negative side by side, North up, East left, with the same intensity scaling, "black" corresponding to a count rate of 0.00017 c/s per pixel after smoothing with a Gaussian kernel of σ=5 pixels (0.125 arcsec). The image scale is 0.025 arcsec/pixel and the images as displayed cover approximately 32×35 arcsec. Filaments are clearly visible in the F150LP image, but not in F165LP. The central region of M87 is visible as diffuse emission to the right in the F150LP image and to a lesser extent in the F165LP image. In addition to extensive filaments throughout the F150LP image, there are a few point sources and a resolved patch of emission in the center of the field, which is the SE synchrotron hotspot of the M87 radio source (Sparks et al. 1992). It has been found that the ACS/SBC is susceptible to redleak (Boffi et al. 2007); however the combination of F165LP and F150LP provides a guard against this. If the wavelength of emission from the filaments were longer than 1700 Å the filaments would be of equal strength in these two images (see Figure 1 which shows the response curves of the two filters which are used in combination with the same detector). Clearly that is not the case. Hence we can conclude with confidence that the emission is not due to redleak.

Figure 3 compares the FUV image to optical data. Images are North up, East left. The left hand panel shows the FUV fields in M87, dominated by the "outer" field as discussed, but with part of the inner field (also GO-11681) included for reference. The underlying galaxy continuum prominent in Figure 2 has been largely removed by scaling and subtracting an elliptical model of M87 derived from optical HST images. The poor subtraction of the optical model in the central region appears to be due to an optical/FUV color gradient in the host galaxy. The subtraction was optimized for the outer field, and has worked well over that region. The right hand panel is an optical HST image of M87 derived from the co-addition of approximately 50 orbits of ACS Wide Field Camera (ACS/WFC) F606W imaging (GO-10543, AR-11283). An elliptical model of the galaxy has been subtracted from the image to reveal a large number of point sources (mostly globular clusters in M87), the base of the famous M87 jet (to the right), the SE lobe synchrotron hotspot (Sparks et al. 1992), some background galaxies, and optical Hα+[NII] filaments which are present within the bandpass of the F606W filter (which is approximately 4700 Å to 7100 Å). The filament system has been studied extensively at higher signal-to-noise using ground based data (Sparks et al. 1993; Sparks et al. 2004) however this new HST image offers a significantly better spatial resolution view of the outer Hα+[NII] system, useful for direct comparison to the FUV data.

The similarity between the FUV filaments and the optical filaments is striking. The filaments in both optical and FUV show the same large scale structure, extending approximately 30 arcsec, width 2 – 3 arcsec, across the outer field, with a hint of continuation into the inner field. They also both show fine structure within the larger scale distribution on length scales down to the image resolution which is about 0.1 arcsec (the optical data are sampled with 0.05 arcsec pixels). The filament fine structure in optical and FUV is spatially coincident to within the ability of the data to discriminate. While subtle differences between the optical and FUV may be present, given the signal-to-noise we have, we cannot be certain of this and deeper images are needed. The data as they stand are consistent with the optical and FUV filaments being identical on all spatial scales with constant flux ratio. These filament complexes are also visible in Chandra X-ray data and are identified and discussed in Sparks et al. (2004).

We used the *stsdas.hst_calib.synphot* synthetic photometry routine package to determine a flux calibration for the optical and FUV emission lines. We took the M87 redshift to be 0.00436, hence a CIV 1549 Å line redshifted wavelength of 1556 Å as the reference wavelength in the *bandpar* routine, which provides the emission line sensitivity of $1.17 \times 10^{-14}$ erg s$^{-1}$cm$^{-2}$ for unit detected countrate. For the optical filaments, the emission is dominated by the sum of [NII] 6548+6584 Å and H$\alpha$ 6563 Å. The average flux ratio of [NII] 6584 Å /H$\alpha$ 6563 Å $\approx 2.52$, and [NII] 6548 Å /[NII] 6584 Å $\approx 1/3$ (Ford & Butcher 1979), so the intensity weighted wavelength of $\lambda \approx 6572$ Å is redshifted to wavelength $\lambda \approx 6601$ Å. The resultant absolute sensitivity of the F606W ACS/WFC combination is $1.584 \times 10^{-16}$ erg s$^{-1}$cm$^{-2}$ for unit countrate. We then generated a mask that isolated the emission regions in the F150LP image, and another that isolated continuum emission from the background galaxy. To generate the emission mask, we used a threshold of $8 \times 10^{-5}$ counts/sec based on the background subtracted F150LP image, resampled and registered onto the optical coordinate system (0.05 arcsec pixels), smoothed with a Gaussian $\sigma=5$ pixels (0.25 arcsec). Some minor adjustments were made by hand to remove noisy areas and point sources. The continuum mask was the complement of the emission mask, further processed to remove point sources and the SE lobe hotspot.

Figure 4 shows the average surface brightness profile in radial bins within the masked region for the CIV and H$\alpha$+[NII] line emission. Included for comparison are the peak surface brightnesses reached within each radial bin to provide an indication of the amplitude of fine structure. The values for the H$\alpha$+[NII] are in good agreement with the data presented in Sparks et al. (2004). The average flux in H$\alpha$+[NII] = $1.43 \times 10^{-15}$ erg s$^{-1}$ cm$^{-2}$ arcsec$^{-2}$, and the average flux in CIV 1549 Å = $0.72 \times 10^{-15}$ erg s$^{-1}$ cm$^{-2}$ arcsec$^{-2}$. Total fluxes are $6.83 \times 10^{-14}$ erg s$^{-1}$ cm$^{-2}$ and $3.43 \times 10^{-14}$ erg s$^{-1}$ cm$^{-2}$ respectively summing over the emission mask area of 47.8 arcsec$^2$, which corresponds to a total power in CIV of $1.1 \times 10^{39}$ erg s$^{-1}$ from this region.

## 3. DISCUSSION

The two most obvious candidates for the origin of the FUV emission are hot gas and stars. Given the strong morphological similarity of the FUV filaments to the H$\alpha$+[NII] filament system, we investigate the hypothesis that FUV filaments are due to line emission from hot gas, specifically CIV 1549 Å which arises from gas at temperatures $\sim 10^5$K. Alternatively, if the FUV emission features are continuum emission, then there is the possibility that they are due to hot UV-emitting stars.

Such stars could potentially be associated with recent star formation within the H$\alpha$+[NII] filaments. The FUV emission is, to the resolution of HST, co-spatial with the optical line filaments, which is of order 0.1 arcsec or around 10 pc. Typical projected velocities within the filament system are (Sparks et al. 1993) $\sim 100$ km s$^{-1}$. A velocity difference of 100 km s$^{-1}$ separates the stars and filaments in about $10^5$ yrs, much less than typical main sequence lifetimes for massive stars of around $10^7$ years. For two systems to remain co-spatial to within 0.1 arcsec over a stellar evolutionary timescale of $\sim 10^7$yr, the relative velocity has to be no more than 1 km s$^{-1}$. A star which forms in a filament responds to the gravitational field of M87, whereas the dynamics of the filament are likely to be determined mainly by more dissipative interactions with local gas (Sparks et al. 1993). At a radius of 30 arcsec (around $R \approx 2.5$ kpc) from the nucleus, the rotational velocity of M87 is $v \approx 400$ km s$^{-1}$ which gives us a measure of the local gravitational field $g = v^2/R$. A star released from rest in such a field moves across the width $w$ of the filament ($w = 10$ pc) in a time of $\sqrt{(2w/g)} = 5 \times 10^5$ yr. Thus, on dynamical grounds, we do not expect any stars, even if formed within the filaments, to currently be within them. Thus, forming and keeping the FUV and Hα+[NII] filaments co-spatial for timescales as long as the stellar main sequence lifetimes is problematic.

In addition, if the FUV is due to hot stars, then the sensitivity of the ACS/SBC is such that individual main sequence stars of early spectral type would be visible. The flux limit corresponds to an FUV luminosity equivalent to the luminosity of an individual B0V star based on Kurucz (1993) models available as a standard library grid within the *stsdas.synphot* distribution, together, for consistency, with stellar parameters from Schmidt-Kaler (1982). Figure 3 illustrates the massive end of a solar metallicity main-sequence with stars of type O3V, O5V, O6V, O8V and B0V easily visible. Later stars are included but are too faint to be seen. Any stars earlier than B0V, mass $M_{min}=17.5$ M$_\odot$, would be clearly visible as individual point sources.

Further, if we suppose that the FUV emission is due to a population of hot stars then we may estimate the total number of stars required, and in particular estimate the number of massive luminous stars that we would expect to be individually detectable. To do this we make use of the stellar population models by Leitherer, Robert & Heckman (1995). For their instantaneous burst model with a lower limit of 1 M$_\odot$, and upper limit of 80 M$_\odot$ and Salpeter type power-law mass distribution with $\alpha = 2.35$, where the IMF of the stellar population is parameterized as

$$\phi(m) = \frac{dN}{dm} = Cm^{-\alpha}, \qquad (1)$$

with $m$ in solar masses. For Leitherer et al.'s canonical mass burst of $10^6$ M$_\odot$ this gives $C = 4.46 \times 10^5$. This burst gives a flux at 1500 Å of $1.9 \times 10^{39}$ erg s$^{-1}$ Å$^{-1}$. At a distance of $D = 16.7$ Mpc this corresponds to a flux of $F_\lambda = 6.0 \times 10^{-14}$ erg s$^{-1}$Å$^{-1}$ cm$^{-2}$. The observed flux is $F_\lambda$(obs) $= 1.26 \times 10^{-16}$ erg s$^{-1}$Å$^{-1}$ cm$^{-2}$ using a unit countrate sensitivity for F150LP of $4.39 \times 10^{-17}$ erg s$^{-1}$Å$^{-1}$ cm$^{-2}$ appropriate for continuum emission, which therefore corresponds to a total burst mass, using this model, of $M(stars) = 2.14 \times 10^3$ M$_\odot$. From the IMF, $\phi$, this would give a total number of individually observable stars with masses $M > M_{min}$ of 14, clearly not a plausible description of the filament morphology. We conclude that the FUV emission is not due to hot young stars and therefore investigate the alternative possibility that it is due to hot gas.

## 4. THEORETICAL CONDUCTION MODEL

If the FUV emission is due to hot gas, then the gas needs to be heated in some manner. We note first that heating the gas by hot stars is ruled out by the above considerations. If we have to rely on the continua from hot stars to heat the gas, then we should *a fortiori* be able to see the stars.

We therefore investigate a simple model in which the presence of cool Hα+[NII] emitting filaments in a surrounding hot X-ray emitting medium implies the likelihood of a range of temperatures at the interface between the two and the possibility of thermal heat flow. We use the approach of Nipoti & Binney (2004) who argued that the filaments currently brightest are those which are on the borderline between evaporation and condensation, and so approximately in a steady state. In this model the filaments are static, with neither evaporation nor condensation occurring. Inflow of heat due to thermal conduction from the ambient medium is exactly balanced by radiation and there is no velocity flow. The competition between evaporation and condensation when a cool body of gas is immersed in a hotter surrounding medium has been studied extensively (e.g. Cowie & McKee 1977; McKee & Cowie 1977; Böhringer & Fabian 1989; McKee & Begelman 1990; Nipoti & Binney 2004). A time dependent computation, starting with a cool filament inserted into hot surroundings will become quasi-steady from inside out, suggesting the inner regions are more likely to be described by the static model, but more work needs to be done for a model extending all the way into the X-ray regime. Hence, while the assumptions of the fiducial model are unlikely to be correct on all scales at all times, they may be a good approximation on small scales, where most of the optical and UV emitting gas resides, and for the most prominent filaments.

For our fiducial model we used the Sutherland & Dopita (1993) cooling function for [Fe/H] = -0.5. Sutherland & Dopita (1993) also give the electron fraction as a function of temperature, which enables us to calculate the total pressure, assumed uniform. We use Spitzer conductivity and fix the inner radius temperature $T_{in} = 10^4$ K and take $dT/dR = 0$ at radius $R_{in}$, so there is no conducted heat flux into the cool body of the filament. Thus we neglect the thermal properties of the cold filament core (Nipoti & Binney 2004). Hence we solve the resulting single second order, ordinary differential equation for temperature $T$ in terms of cylindrical radius $R$ using standard methods (e.g. Press et al. 1992). Armed with the temperature structure, we obtain line emissivities using 'CLOUDY' (Ferland et al. 1998; http://www.nublado.org). Figure 1 shows the resulting UV spectrum for a model with $R_{in}$=10 pc, n=0.1 cm$^{-3}$, T=5×10$^6$K (0.7×10$^{-10}$dynes cm$^{-2}$) and standard Spitzer conductivity. For such a filament the flux emitted per unit length in the CIV 1549 Å line is 4.5x10$^{14}$($R_{in}$/10 pc)$^2$ erg s$^{-1}$cm$^{-1}$. The actual pressure in M87 is uncertain, but likely in the range 6-25×10$^{-10}$dynes cm$^{-2}$ (Young, Wilson & Mundell 2002; Sparks et al. 2004) and the model is sensitive to the pressure cubed. The total length of the observed filament system is approximately 2.5 kpc in projection. Thus the model would predict a total CIV flux of ~5x10$^{39}$ erg s$^{-1}$ for a pressure 8×10$^{-10}$dynes cm$^{-2}$ for the North-East quadrant (Young et al. 2002). This is consistent with the observed value of 1.1x10$^{39}$ erg s$^{-1}$ given significant sensitivities to poorly determined parameters (inner radius; pressure; geometry; conduction suppression factor) and, while not a definitive model, suffices to show that thermal conduction provides a plausible explanation for the interface.

5. CONCLUSIONS

We have presented FUV images that reveal filaments in the galaxy M87 at the center of the Virgo cluster. The filaments are remarkably similar in morphology to, and coincident with, the well-known low-excitation optical emission filaments in this galaxy. There are also known to be hot thermal X-ray filaments at the same location. We compare the possibilities that the filaments are continuum emission from hot stars, versus line emission from the CIV 1549 Å resonance line. We conclude that the filaments are due to CIV line emission, which in turn implies the presence of gas at temperature ~$10^5$K. This temperature is in the interesting transition region between hot coronal X-ray emitted gas at ~$10^7$K and cooler optical line emitting gas at ~$10^4$K. We show that the strength of the intermediate temperature line emission is consistent with a model of a thermal conduction interface. The results have profound implications for models in which low excitation filaments are photoionized by hot young stars (we do not see any). If thermal conduction proves to offer the correct description of the interface, then transport processes in the hot interstellar medium of cool core galaxy clusters are present and the energy sources within the galaxy are available to replenish losses due to radiative processes.

ACKNOWLEDGEMENTS

Support for this work was provided by NASA through grants GO-11681 and AR-11283 from the Space Telescope Science Institute, which is operated by the Association of Universities for Research in Astronomy, Inc., under NASA contract NAS5-26555.

REFERENCES

Blakeslee, J.P., Jordán, A., Mei, S., Côté, P, Ferrarese, L., Infante, L., Peng, E.W., Tonry, J.L., West, M.J., 2009, ApJ, 694, 556
Boffi, F.R., et al. 2007, "ACS Instrument Handbook", Version 8.0, (Baltimore: STScI)
Böhringer, H., Fabian A.C., 1989, MNRAS, 237, 1147
Braine, J., Wyrowski, F., Radford, S.J.E., Henkel, C., Lesch, H, 1995, A& A, 293, 315
Conselice, C.J., Gallagher, J.S., Wyse, R.F.G., 2001, AJ, 122, 2281
Cowie, L.L., Fabian, A.C., Nulsen, P.E.J., 1980, MNRAS, 191, 399
Cowie, L.L., McKee, C.F., 1977, ApJ., 211, 135
Crawford, C.S., Hatch, N.A., Fabian, A.C., Sanders, J.S., 2005, MNRAS, 363, 216
Donahue, M., Voit, G.M., 1991, ApJ, 381, 361
Fabian, A.C., Nulsen, P.E.J., 1977, MNRAS, 180, 479
Fabian, A.C., Nulsen, P.E.J., Canizares, C.R., 1984, Nature, 310, 733
Fabian, A.C., Sanders, J.S., Crawford, C.S., Concelice, C.J., Gallagher, J.S., Wyse, R.F.G., 2003, MNRAS, 344, L48
Ferland, G.J., Korista, K.T., Verner, D.A., Ferguson, J.W., Kingdon, J.B., Verner, E.M., 1998, PASP, 110, 761
Ford, H.C., Butcher, H., 1979, ApJS, 41, 147
Hatch, N.A., Crawford, C.S., Johnstone, R.M., Fabian, A.C., 2006, MNRAS, 367, 433
Hatch, N.A., Crawford, C.S., Fabian, A.C., 2007, MNRAS, 380, 33
Heckman, T.M., Baum, S.A., van Breugel, W.J.M., McCarthy, P., 1989, ApJ, 338, 48
Kurucz, R.L., 1993, from the Kurucz database, Goddard Space Flight Center, CD-ROM #13 implemented within *synphot* (STScI: Baltimore)
Leitherer, C., Robert, C., Heckman, T.M., 1995, ApJ, 503, 646
McKee, C.F., Begelman, M.C., 1990, ApJ, 358, 392
McKee, C.F., Cowie, L.L., 1977, ApJ, 215, 213
Nipoti, C., Binney, J., 2004, MNRAS, 349, 1509
Press, W.H., Teukolsky, S.A., Vettering, W.T., Flannery, B.P., 1992, Numerical Recipes, Cambridge University Press
Revaz, Y., Combes, F., Salomé, P., 2007, A&A, 477, L33
Rubin, V.C., Oort, J.H., Ford, W.K., Peterson, C.J., 1977, ApJ, 211, 693
Sabra, B.M., Shields, J.C., Fillipenko, A.V., 2000, ApJ, 545, 157
Sabra, B.M., Shields, J.C., Ho, L., Barth, A.J., Filippenko, A.V. 2003, ApJ, 584, 164
Schmidt-Kaler, Th., 1982, in Landolt-Börnstein, Group IV, 2b, Stars and Star Clusters; Eds. K. Schaifers & H.H. Voigt
Sparks, W.B., 1992, ApJ, 399, 66
Sparks, W.B., Macchetto, F., Golombek, D., 1989, 345, 153
Sparks, W.B., Fraix-Burnet, D., Macchetto, F, Owen, F.N., 1992, Nature, 355, 804
Sparks, W.B., Ford, H.C., Kinney, A.L. 1993 ApJ, 413, 531
Sparks, W.B., Donahue, M., Jordán, A., Ferrarese, L., Coté, P., 2004, ApJ, 607, 294
Sutherland, R.S., Dopita, M.A., 1993, ApJS, 88, 253
Voit, G.M., Donahue, M., 1997, ApJ, 486, 242
Yagi, M., Komiyama, Y., Yoshida, M., Furusawa, H., Kashikawa, N., Koyama, Y., Okamura, S., 2007, ApJ, 660, 1209
Young, A.J., Wilson, A.S., Mundell, C.G., 2002, ApJ, 579, 560

FIGURES:

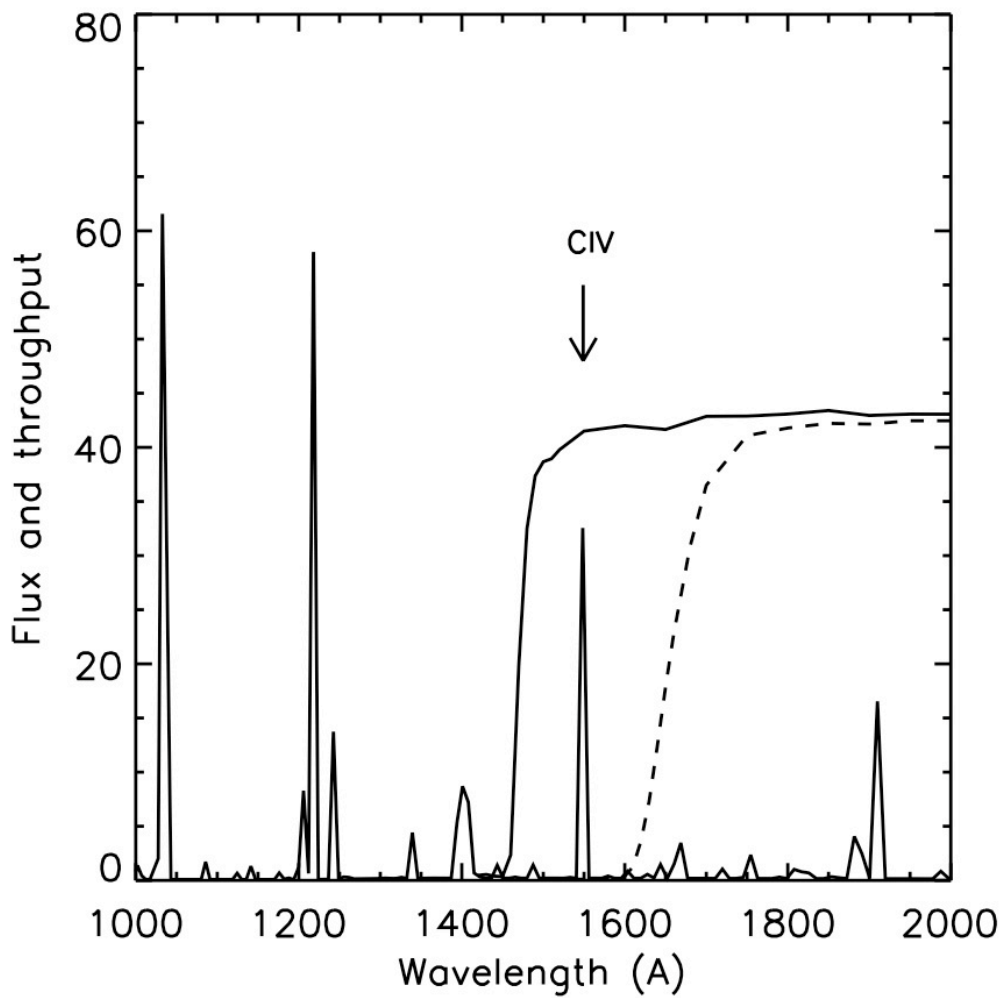

Figure 1. Fiducial model FUV emission line conduction spectrum as described in text. Overlaid are the throughput curves for the F150LP and F165LP filters of the ACS/SBC, showing how they bracket CIV 1549. A difference image between the two is equivalent to a CIV image.

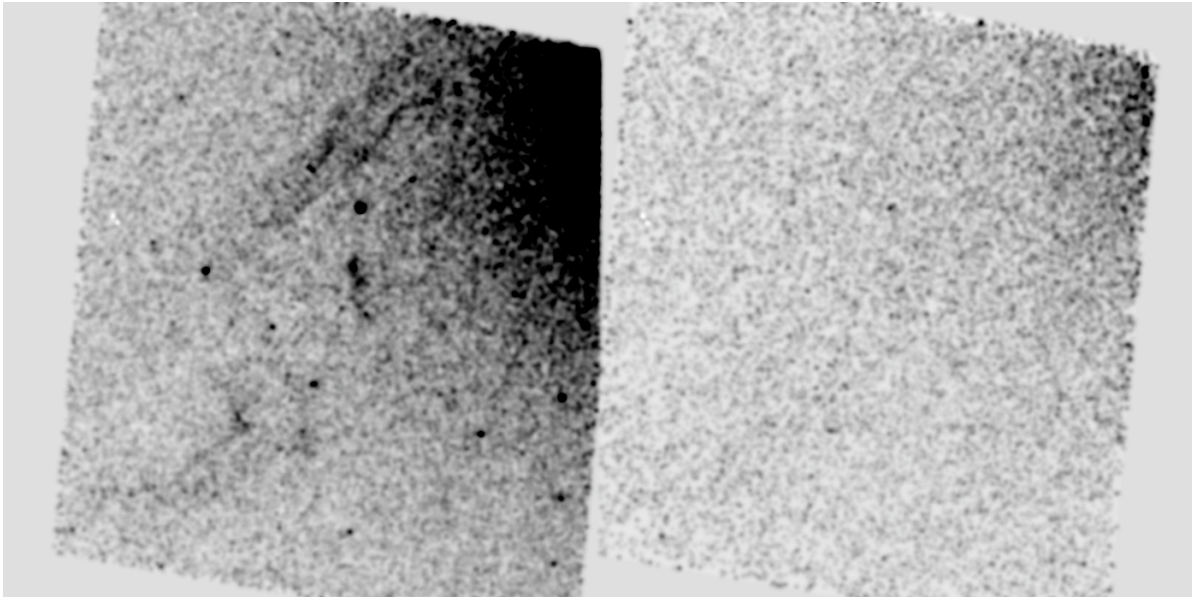

Figure 2. F150LP (left) and F165LP (right) smoothed count rate images using the same intensity scaling, North up, East left, shown on a negative intensity scale with "black" corresponding to 0.00017 counts s$^{-1}$ pixel$^{-1}$. The images are approximately 32×35 arcsec. Filaments are clearly visible in the F150LP image, but not in F165LP, demonstrating that the emission is not due to redleak.

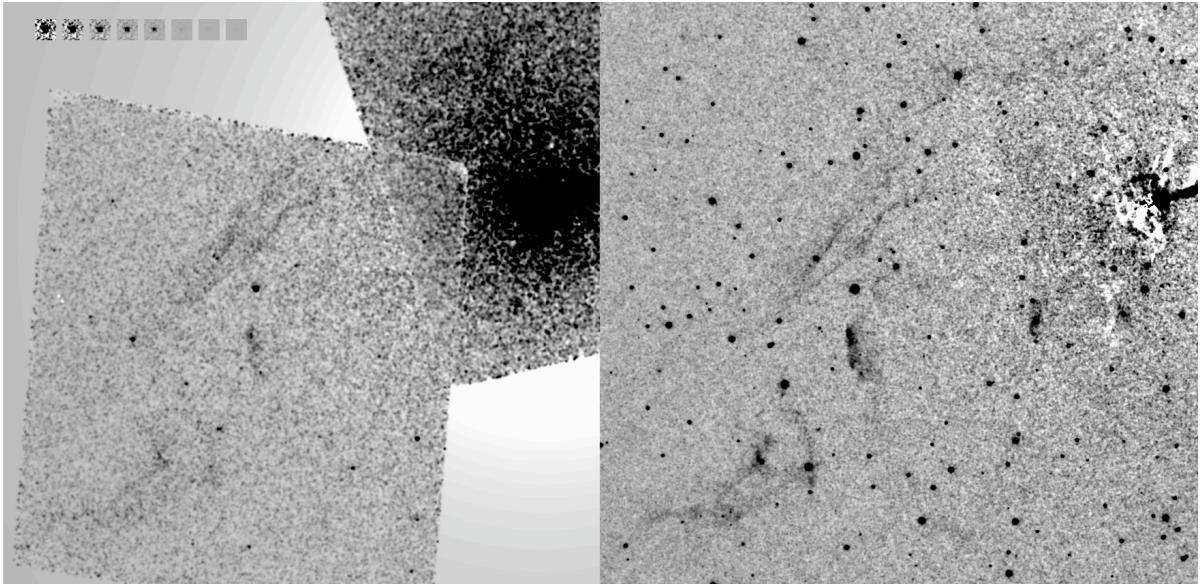

Figure 3. A comparison of the FUV F150LP image (left) to an ACS/WFC F606W optical image (right) shown with a negative intensity scale. Images are North up, East left. The row of artificial point sources inserted upper left shows individual upper main sequence stars as they would appear in this image, left to right O3V, O5V, O6V, O8V, B0V. Later types are included but cease to be visible beyond B3V. Fluxes were derived using Kurucz (1993).

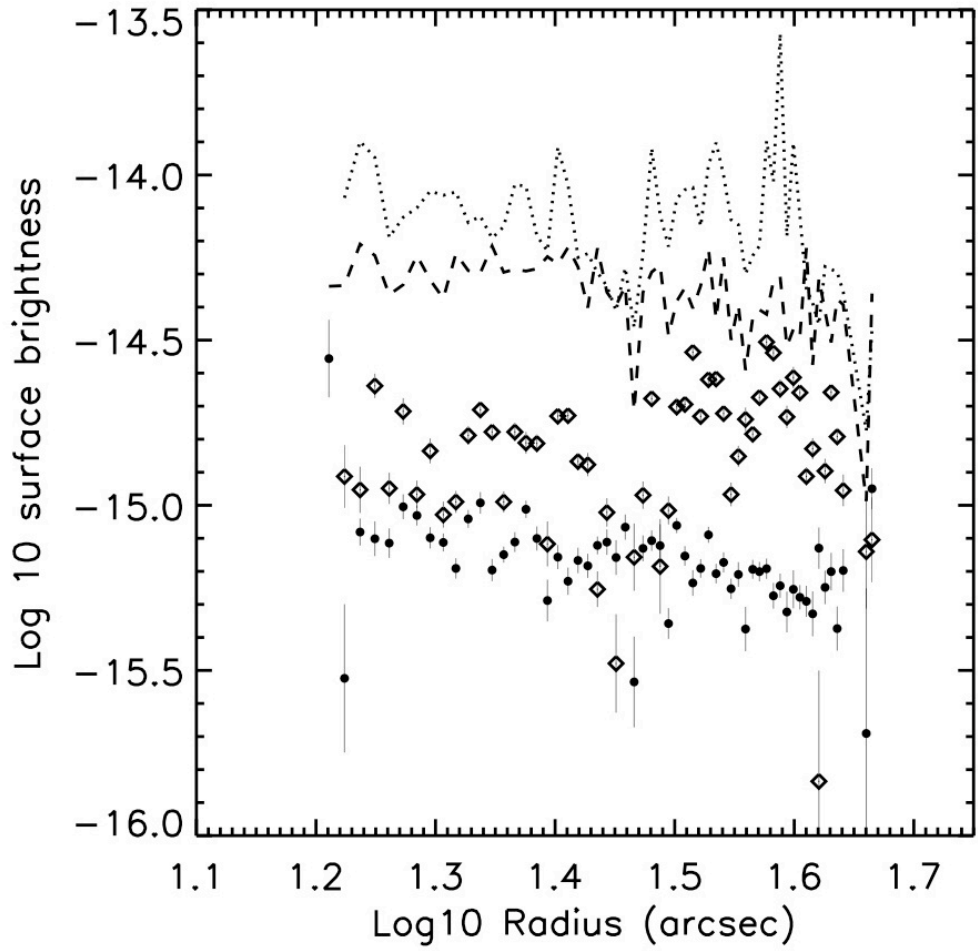

Figure 4. The surface brightness profiles in radial bins for CIV and Hα+[NII] line emission. The solid symbols are CIV surface brightness and the open symbols, Hα+[NII]. The dashed line above is the peak CIV surface brightness and the dotted line, the peak Hα+[NII] surface brightness. Units are erg s$^{-1}$ cm$^{-2}$ arcsec$^{-2}$.